\documentclass[letterpaper]{article} 
\usepackage{aaai2026}  
\usepackage{times}  
\usepackage{helvet}  
\usepackage{courier}  
\usepackage[hyphens]{url}  
\usepackage{graphicx} 
\usepackage{natbib}  
\usepackage{caption} 
\usepackage{algorithm}
\usepackage{algorithmic}
\usepackage{amsmath}
\usepackage{booktabs}
\usepackage{multirow}
\usepackage{adjustbox}

\usepackage{newfloat}
\usepackage{listings}
\DeclareCaptionStyle{ruled}{labelfont=normalfont,labelsep=colon,strut=off} 
\floatstyle{ruled}
\newfloat{listing}{tb}{lst}{}
\floatname{listing}{Listing}
\title{DiffHash: Text-Guided Targeted Attack via Diffusion Models against Deep Hashing Image Retrieval}
\author{
    Zechao Liu\textsuperscript{\rm 1},
    Zheng Zhou\textsuperscript{\rm 1},
    Xiangkun Chen\textsuperscript{\rm 1},
    Tao Liang\textsuperscript{\rm 1},
    Dapeng Lang\textsuperscript{\rm 1}
}

\affiliations{
    \textsuperscript{\rm 1}Harbin Engineering University, Harbin  150001, China\\
   \{liuzechao, zhouzheng, chenxiangkun, liangtao98, langdapeng\}@hrbeu.edu.cn
}

\usepackage{bibentry}

\begin{document}

\maketitle

\begin{abstract}
Deep hashing models have been widely adopted to tackle the challenges of large-scale image retrieval. However, these approaches face serious security risks due to their vulnerability to adversarial examples. Despite the increasing exploration of targeted attacks on deep hashing models, existing approaches still suffer from a lack of multimodal guidance, reliance on labeling information and dependence on pixel-level operations for attacks. To address these limitations, we proposed DiffHash, a novel diffusion-based targeted attack for deep hashing. Unlike traditional pixel-based attacks that directly modify specific pixels and lack multimodal guidance, our approach focuses on optimizing the latent representations of images, guided by text information generated by a Large Language Model (LLM) for the target image. Furthermore, we designed a multi-space hash alignment network to align the high-dimension image space and text space to the low-dimension binary hash space. During reconstruction, we also incorporated text-guided attention mechanisms to refine adversarial examples, ensuring them aligned with the target semantics while maintaining visual plausibility. Extensive experiments have demonstrated that our method outperforms state-of-the-art (SOTA) targeted attack methods, achieving better black-box transferability and offering more excellent stability across datasets.
\end{abstract}


\section{Introduction}
The exponential growth of multimedia data on the Internet has posed significant challenges for information retrieval systems. As a solution to information retrieval, hashing \cite{wang2017survey} transforms high-dimensional data to binary codes, enabling efficient large-scale search. This approach offers substantial benefits in storage efficiency and retrieval performance. The rise of deep learning has enhanced traditional hashing methods. Leveraging the powerful feature representation capabilities of deep learning, deep hashing models utilize deep neural networks (DNNs) for automatic feature extraction and have achieved significant success.

The DNN-based deep hashing models, with their powerful feature learning capabilities, have quickly surpassed traditional hashing methods that rely on handcrafted features. However, the deep hashing models inherit the vulnerability of the DNNs, making them susceptible to adversarial examples. Despite much research has focused on the security of DNNs like \cite{FGSM, IFGSM, PGD, deepfool, advpatch}, fewer studies address the security problems of the deep hashing models. Due to the discretization of deep hashing models, it is more challenging to generate adversarial examples than classifiers or detectors.  

Existing attacks on deep hashing models could be divided into targeted attacks \cite{bai2020targeted,wang2021prototype,wang2023targeted,wang2021targeted,meng2024targeted,zhao2023precise,tang2024once} and untargeted attacks \cite{saa,cgat}. In this paper, we focuses on targeted attacks, which are more malicious and more challenging to mitigate. Taking online retailers for example, unscrupulous merchants may use targeted attacks on their competitors' products to gain more attention to their own goods, which could cause malignant competition. Compared to the untargeted attack, the targeted attack requires the adversarial examples to be mapped into the preset target hashing codes, which significantly increases the complexity of generating adversarial examples. Although existing methods against deep hashing models have achieved remarkable performance in targeted attacks, they still have several limitations: 1) General methods that rely on structured representation codes and labels often face challenges when dealing with imbalanced datasets and limited label availability for both target and original images. They also lack multimodal semantic guidance, which is vital for effective real-world attacks. 2) Deep hashing models introduce additional complexities for targeted attacks: the non-differentiable sign function causes gradient blockage, while quantization errors arising from binarization lead to reduced semantic accuracy. 3) Most available methods require extensive training for the final attack, which results in expensive resource costs.

To address existing challenges and inspired by recent diffusion-based attacks \cite{dai2025advdiff,chen2024diffusion,xue2023diffusion}, we propose DiffHash, a novel targeted adversarial attack leveraging pre-trained diffusion models against deep hashing. 
The main contributions are summarized as follows: 

\begin{itemize}

    \item We propose a novel targeted attack method against deep hashing models. To the best of our knowledge, we are the first to transfer targeted attacks into the latent space of diffusion models to address the gradient blockage and quantization errors in deep hashing. Our latent-space approach enhances robustness and improves attack transferability across deep hashing architectures.
\item We leverage LLM-based text representations to guide targeted attacks rather than labels and representation codes. Technically, We propose a multi-space hash alignment operation to unify text and image representations in a shared space, and introduce a network HashAlignNet (HAN) to facilitate this alignment. 
    \item Extensive experiments have demonstrated that the generated adversarial examples are more effective and transferable compared to SOTA targeted attack methods. 
\end{itemize}

\begin{figure*}[!htbp]
  \centering
  \includegraphics[width=1 \linewidth ]{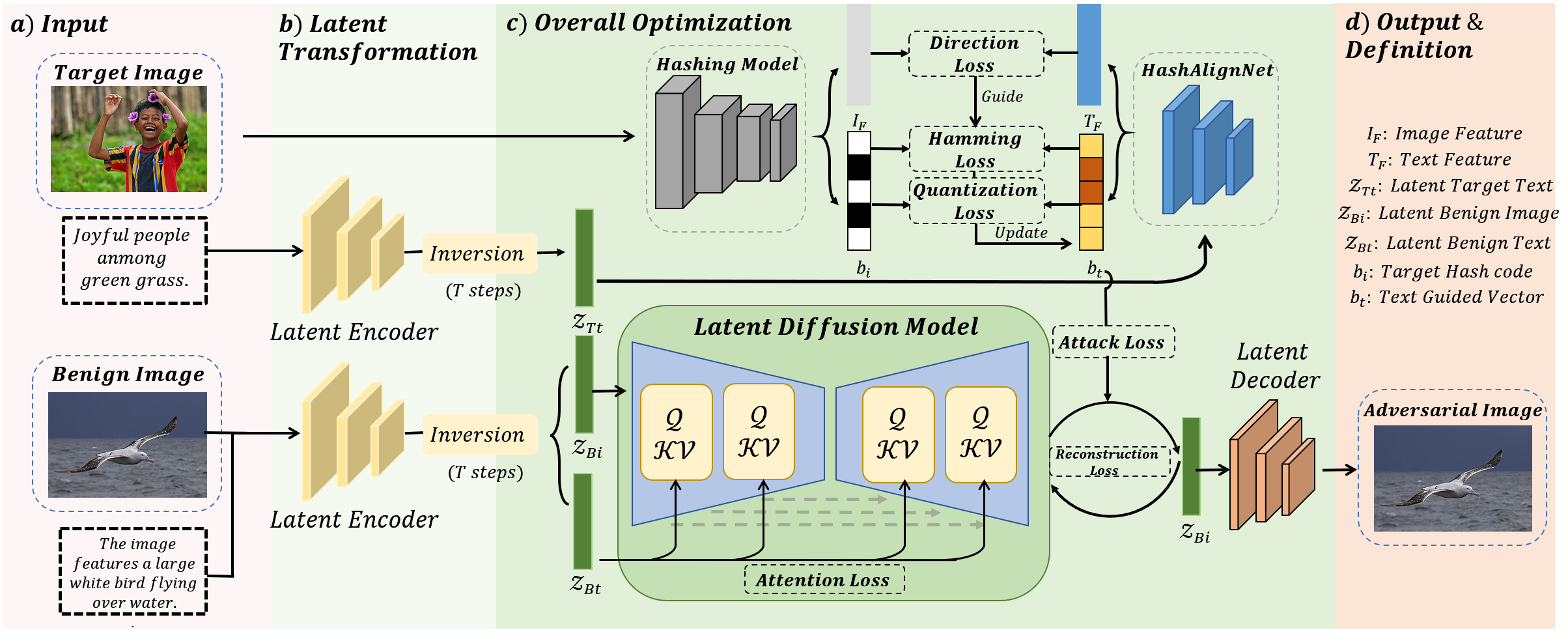}
  \caption{The framework utilizes a latent diffusion model to optimize adversarial examples. Target and benign images with corresponding textual descriptions are encoded into latent representations. They are iteratively refined using multiple loss functions. The optimized latent representations are then decoded to generate adversarial images that effectively achieve targeted attacks on deep hashing models. }
\label{figure1}
\end{figure*}
\section{Related Work}

\subsection{Deep Hashing Based Image Retrieval}
Due to its storage efficiency and retrieval speed, the deep hashing models have become a popular solution for large-scale similarity searches in multimedia applications. Regarding the evolution of the hashing methods, the early techniques were based on hand-crafted approaches, such as Locality-Sensitive Hashing (LSH) \cite{datar2004locality}, which focuses on preserving pairwise similarity between data points. With the fast advancement of deep learning, various deep hashing models have been proposed for image retrieval, including HashNet \cite{cao2017hashnet}, DPSH \cite{li2016feature}, DCH \cite{cao2018deep}, DAPH \cite{shen2017deep}, DPH \cite{cao2018deep11}, and CSQ \cite{yuan2020central}.

\subsection{Adversarial Attacks}
Since Szegedy et al. \cite{FGSM} discovered that small perturbations on images can deceive DNNs and cause misclassifications, adversarial attacks have gained significant attention. As adversarial attack methods continue to evolve, more and more researchers begin to explore the vulnerability of the deep hashing models. Attacks on the deep hashing models can be categorized into two types: targeted attacks and untargeted attacks. Untargeted attack methods include HAG\cite{yang2018adversarial}, AACH \cite{li2021adversarial}, SAA\cite{saa}, CGAT\cite{cgat}. Targeted attack methods can be divided into two main categories: (1) methods that rely on label-based representation codes and generative models, including GAN-based ProS-GAN \cite{wang2021prototype}, THA \cite{wang2021targeted}, TUA \cite{meng2024targeted}, and Diffusion-based HUANG \cite{huang2025huang}; and (2) non-generative methods, such as P2P \cite{bai2020targeted}, DHTA \cite{bai2020targeted}, and PTA \cite{zhao2023precise}. Among these, PTA is the only gradient-simulating method that does not rely on representation codes or labels.

\subsection{Diffusion Models and Latent Space}
Diffusion models are generative models that progressively denoise random noise to generate high-quality data, offering greater training stability and superior sample quality compared to GANs \cite{goodfellow2014generative} and VAEs \cite{kingma2013auto}. It is widely used in image inpainting \cite{li2022mat,xie2023smartbrush}, super-resolution \cite{saharia2022image}, and style transfer \cite{sun2024diffam}. To address the high computational cost of traditional pixel-space diffusion models like DDPM, Latent Diffusion Models (LDMs) such as Stable Diffusion \cite{rombach2022high} operate in a compact latent space, significantly reducing complexity and enhancing versatility. 

\section{Method}
In this section, we will introduce our proposed DiffHash targeted attack method, the overall framework is depicted in Figure \ref{figure1}. DiffHash first encodes all inputs into the latent space of the diffusion model with a \textit{Latent Encoder}. Then the whole process is divided into two stages: in the alignment stage, the encoded latent features are aligned across multiple spaces through the \textit{HashAlignNet}. Subsequently, in the attack stage, the \textit{text guided vector} is employed to optimize the \textit{latent benign image}. Finally, the \textit{Latent Decoder} decodes the optimized \textit{latent benign image} into an \textit{adversarial image}.

\subsection{Problem Formulation} \label{3.1}

Given a dataset $\mathcal{D}={\{ (x_i,y_i)\}}_{i=1}^{N}$, let $N$ denotes the total number of the images, $x_i$ represents the $i$-th image and $y_i = [y_{i1},y_{i2},...,y_{iC}]\in \{0,1\}^C$ corresponds to a multi-label vector labeled with $C$ classes, where $y_{in}=1$ indicates the $x_i$ belongs to the $n$-th class.
The semantic similarities between any two images $S_{i,j}$ are defined as
\begin{equation}
    S_{i,j} = 
    \begin{cases}
        1, &  y_{i} \cap y_{j} \neq \emptyset \\
        0, &  otherwise
    \end{cases}
\end{equation}
We consider two images similar if they share at least one common label. A deep hashing model $\mathcal{H}$ maps data (e.g. images) to binary codes and computes the similarities within the database. Generally, given the input image $x_i$, the deep hashing model will map it to a hash code: 
\begin{equation}
    b(x_i) = sign(\mathcal{H}(x_i)) ,    b(x_i) \in \{-1, 1\}^k
\end{equation}
where $sign(\cdot)$ represents the sign function, producing binary values in $\{-1,1\}$, and $k$ is the length of the hash code.
To evaluate the similarity between two images $x_i$ and $x_j$, their respective hash codes $b(x_i)$ and $b(x_j)$ are compared using Hamming distance:
\begin{equation}
    d_{H}(x_i,x_j) =  \frac{1}{2} \sum_{l=1}^{k}|b_{l}(x_i)-b_{l}(x_j)|
\end{equation}
where $b_l(x_i)$ and $b_l(x_j)$ represent $l$-th bit of hash codes for $x_i$ and $x_j$. 

\subsection{Multi-Space Hash Alignment}
\begin{figure}[!htbp]
  \centering
  \includegraphics[width=1 \linewidth ]{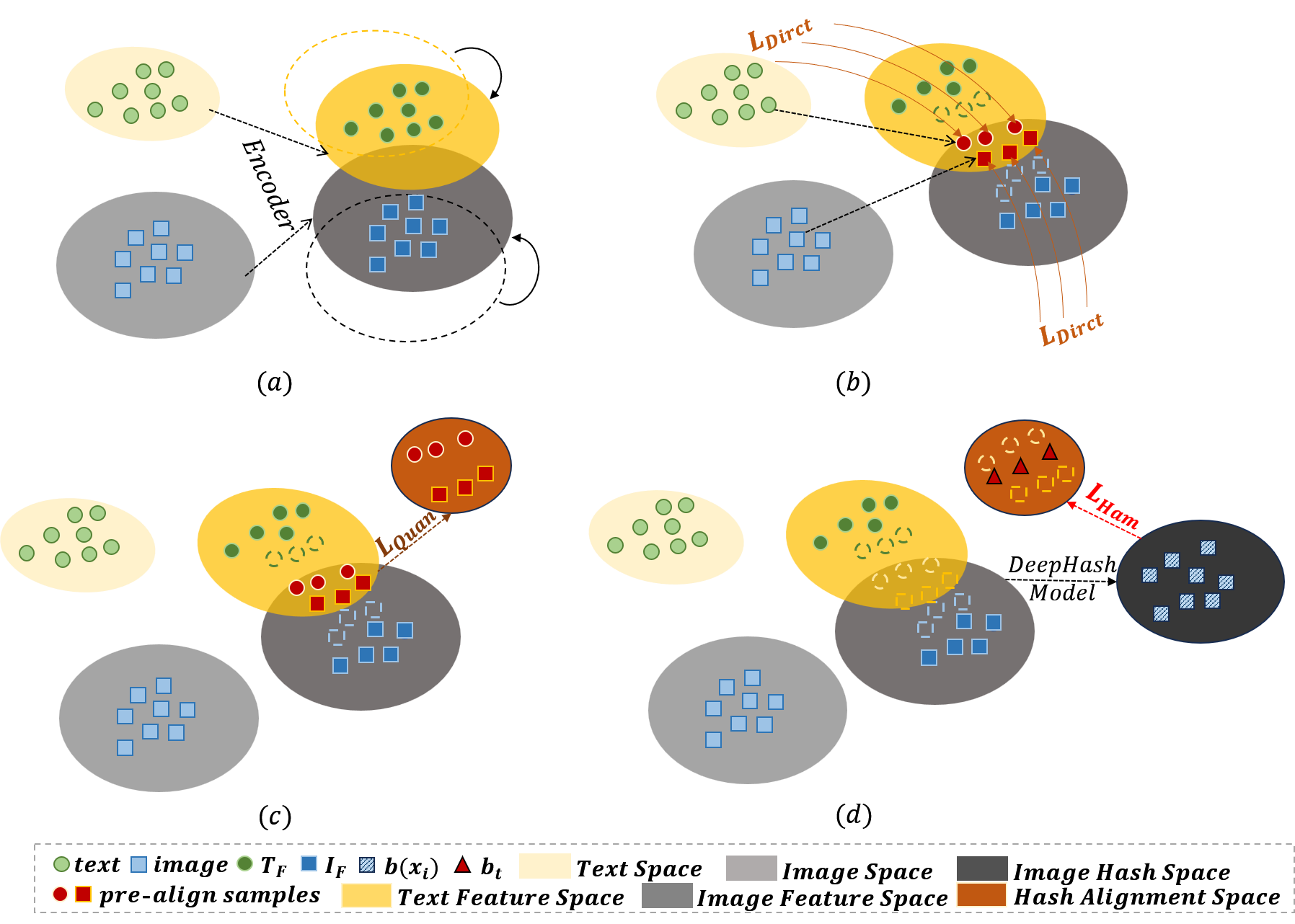}
  \caption{The process of Multi-Space Hash Alignment: (a) text and image samples are first encoded into their feature spaces, (b) then features are brought closer using $L_{\rm Direct}$, (c) these pre-algin samples are quantized into binary codes via  $L_{\rm Quan}$, (d) finally the $L_{\rm Ham}$ further aligns text and image hash codes.}
  \label{fig:figure2}
\end{figure}

Existing attacks on deep hashing models mostly rely on label information within single-image modalities, restricting their adaptability and effectiveness. We propose a novel targeted attack method that leverages LLM-based textual representations to guide adversarial attacks. To bridge the inherent gap between high-dimensional image and text space, we introduce a Multi-Space Hash Alignment approach which implemented via our HashAlignNet (HAN). HAN comprises three fully connected layers ($1024 \rightarrow 512 \rightarrow 256$). As illustrated in Figure \ref{fig:figure2}, HAN efficiently maps text features into a shared hash space. The input latent target text $z_{Tt}$, typically derived from a pre-trained text encoder such as Stable Diffusion \cite{rombach2022high}, is projected through HAN into the hash space, yielding text feature $T_F$ and text guided vector $b_t$. 

To ensure the hash representations maintain semantic alignment, we introduce the direction consistency loss $\mathcal{L}_{Direct}$ as: 
\begin{equation}
    \mathcal{L}_{Direct} = 1- \frac{1}{N} \sum_{i=1}^{N}\frac{ T_{F}^{(i)} \cdot I_{F}^{(i)}}{\left \| T_{F}^{(i)} \right \|_2 \cdot \left \| I_{F}^{(i)} \right \|_2}
\end{equation}

where we use cosine similarity to control the direction of the alignment, minimize the cosine distance between $T_{F}^{(i)}$ and $I_{F}^{(i)}$. The $I_{F}^{(i)},T_{F}^{(i)}$ is the $i$-th image feature and text feature extracted from the target hash model and HAN. $\left \| \cdot \right \|_{2}$ calculates the $L_2$ norm of $I_{F}^{(i)},T_{F}^{(i)}$.

However, $b_{t}^{(i)}$ may inadvertently exceed the binary constraints within the range of -1 to 1 during optimization. To address this issue, we introduce a quantization loss $\mathcal{L}_{Quan}$ as:
\begin{equation}
\mathcal{L}_{\rm Quan}
=\frac{1}{N}\sum_{i=1}^N
\big\lVert\,\lvert b_t^{(i)}\rvert - \mathbf{1}\big\rVert_{2}
\end{equation}
The text guided vector $b_{t}^{(i)}$ is generated from $i$-th textual input by HAN. This loss discourages deviations in the $L_2$ norm of the textual hash vectors, ensuring that the hash codes remain normalized. 

To effectively align textual and visual modalities for robust and precise attacks, we further introduce hamming loss function $\mathcal{L}_{Ham}$ :
\begin{equation}
    \mathcal{L}_{Ham} =\frac{1}{N} \sum_{i=1}^{N} \left \| b_{t}^{(i)}-b(x_i)\right \|_1
\end{equation}
where $b(x_i) $ denotes the corresponding images' hash codes processed by the deep hash model. $\left \| \cdot \right \|_{1}$ calculates the $L_1$ norm between $b_{t}^{(i)}$ and $b(x_i)$, thereby encouraging their convergence in the shared hash space. 
The total loss of hash alignment is defined as:
\begin{equation}
    \mathcal{L}_{Align}=\alpha \mathcal{L}_{Direct} + \beta \mathcal{L}_{Quan} + \gamma \mathcal{L}_{Ham}
\end{equation}
where the $\alpha$, $\beta$, $\gamma$ are the hyperparameters that weigh the significance of the different losses respectively. 
\subsection{Text-Guided Latent Targeted Attack}
In this section, we propose a text-guided targeted attack that operates in the latent space of the diffusion model. Unlike pixel-level attacks like HUANG \cite{huang2025huang}, our approach overcomes two key challenges specific to deep hashing models: (1) the gradient blockage caused by the non-differentiable binarization sign function, (2) quantization errors arising from mapping continuous vectors into discrete binary codes.

Inspired by recent diffusion editing approaches \cite{couairon2023diffedit,mokady2023null}, we leverage the DDIM Inversion \cite{songdenoising} to shift the optimization into a latent space. Enabling precise alignment of image latent representations with target semantics. At first, we reverse the benign image into the diffusion latent space:
\begin{equation}
    z_{Bi}^{0} = Inverse(x_{i}^{t-1}) = \begin{matrix} \underbrace{Inverse \circ  \ldots \circ Inverse(x_{i}^{0})} \\ t
\end{matrix}
\end{equation}
where $Inverse(\cdot)$ denotes the DDIM Inversion operation, we apply this operation for several timesteps let $x_{i}^{0}$(the benign image) to its latent representation $z_{Bi}^{0}$. 

\begin{figure}[!htbp]
  \centering
  \includegraphics[width=1 \linewidth ]{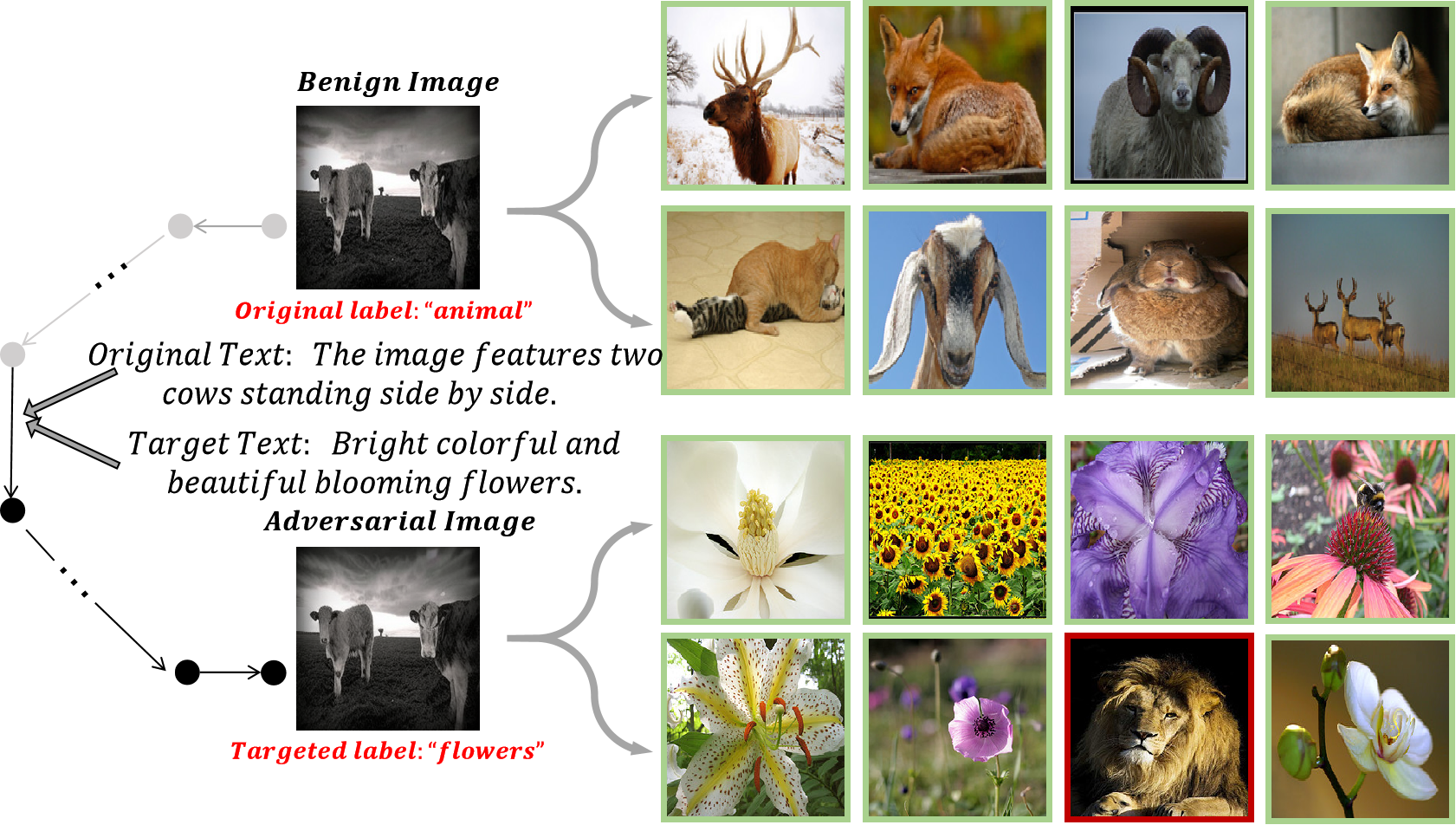}
  \caption{Illustration of the top-8 retrieval results with DiffHash. The adversarial image is generated by the guidance of the targeted text. Images in green/red boxes are the results related/unrelated to the target label.}
\label{fig:figure3}
\end{figure}

To overcome the gradient blockage limitation, we aim to directly perturb the latent representation $z_{Bi}^{0}$ in $T$ steps as:
\begin{equation}
\begin{aligned}
&\mathcal{L}_{Distance} = \sum_{t=0}^{T} \mathcal{J}(b_{t}, b(z_{Bi}^{t}); \mathcal{H})  
\end{aligned}
\end{equation}

where $\mathcal{J}$ is the cross-entropy loss. $z_{Bi}^{t}$ denotes the latent vector in the $t$-th adversarial optimization update. And the $b_{t}$ is the text-guided vector, which is obtained from HashAlignNet. 
To better guide the optimization and manage the quantization errors of the perturbation, we incorporate directional constraints, the loss $\mathcal{L}_{Path}$ are defined as:
\begin{equation}
\begin{aligned}
    \mathcal{L}_{Path} 
    &=\sum_{t=0}^{T}  \left \| b_{t}-b(z_{Bi}^{t})\right \|_{1} \\
    & + \sum_{t=0}^{T}  max(0, M_{margin} - (b_{t}  \cdot b(z_{Bi}^{})))
\end{aligned}
\end{equation}
where $max(0,\cdot)$ ensures that when the dot product $b_{t}  \cdot b(z_{Bi}^{t})$ is less than the threshold $M_{margin}$, the penalty will be triggered to constraint the similarity of the $b_{t}$ and $b(z_{Bi}^{t})$. Finally, the total attack loss combines the $\mathcal{L}_{Distance}$ and $\mathcal{L}_{Path}$ to optimize the adversarial perturbation in the latent space jointly:
\begin{equation}
    \mathcal{L}_{Attack} =  \mathcal{L}_{Distance} +  \mathcal{L}_{Path}
\end{equation}
Figure \ref{fig:figure3} shows an example of retrieval results, comparing a benign image with its adversarial example generated by our method. 
\begin{table*}[!htbp]

	\begin{center}
			\begin{tabular}{l c ccc lllccc} 
				\toprule
				\multirow{2}{*}{Method} & \multirow{2}{*}{Metric} & \multicolumn{3}{c}{NUS-WIDE}&  \multicolumn{3}{c}{FLICKR-25K}&\multicolumn{3}{c}{MS-COCO} \\ 
				\cmidrule(lr){3-5} \cmidrule(lr){6-8} \cmidrule(lr){9-11} &                 & \textbf{16 bits}& \textbf{32 bits}& \textbf{64 bits}&  \textbf{16 bits}& \textbf{32 bits}& \textbf{64 bits}&\textbf{16 bits}& \textbf{32 bits}& \textbf{64 bits}\\ 
				\midrule
				Original         & t-MAP    & 27.38 & 26.70 & 25.52 &  43.52 & 42.82 & 42.82 
&24.01 & 21.69 & 20.15 \\
				P2P& t-MAP    & 39.39 & 35.88 & 42.49 &  59.86 & 57.15& 52.41
&30.97 & 29.56 & 28.87 \\
				DHTA& t-MAP    & 39.98 & 36.13 & 43.34 &  50.93& 52.36& 52.13
&30.82 & 29.50 & 29.09 \\
				ProS-GAN& t-MAP    & 67.01 & 74.31& 69.37&  63.41& 60.99 & 57.60
&34.67 & \underline{55.71} & 50.21 \\
				THA& t-MAP    & 60.65 & \underline{75.77}& \underline{74.63}&  \textbf{66.69} & 60.31& \underline{60.44}
&38.95 & 51.71 & 32.51 \\
				PTA& t-MAP    & \textbf{71.31}& 72.59 & 71.64 &  47.02 & 55.61 & 54.23 
&\underline{57.49} & 52.76 & \underline{60.24}\\
 HUANG& t-MAP& 57.12& 58.61& 63.55& 60.22& \underline{61.15}& 59.43& 48.16& 49.36&50.44\\\hline
				DiffHash (Ours)  & t-MAP    & \underline{69.13} & \textbf{79.97} & \textbf{77.62}&  \underline{65.32} & \textbf{62.67} & \textbf{61.33}&\textbf{62.46} & \textbf{59.42} & \textbf{61.49}\\
				\bottomrule
			\end{tabular}
		
	\end{center}
		\caption{t-MAP@5000 of different targeted adversarial attack methods with 16,32 and 64 hash bits on three datasets. The best results are highlighted in bold, while the second-best results are underlined.}
    \label{tab:results1}
\end{table*}
\subsection{Optimization and Reconstruction}
During denoising iterations, attention maps are aggregated to analyze the model's focus on image regions relevant to the original text. By leveraging textual guidance derived from the original image, we constrain deviations in attention distribution through variance minimization. This approach ensures semantic coherence and maintains consistent attention to textually relevant regions throughout the optimization process. This is mathematically represented as: 
\begin{equation}
    \mathcal{L}_{\mathrm{Attention}}=\frac{1}{H\times W\times S}\sum_{s=1}^{S}\sum_{h=1}^{H}\sum_{w=1}^{W}\left(\mathcal{A}_{h,w}^{s}(z_{Bt})-\mu_{s}\right)^{2}
\end{equation}
where $\mathcal{A}_{h,w}^{s}$ denotes the attention value at location $(h, w)$ for the $s$-th attention head, $S$ is the number of attention heads, $H$ and $W$ are the spatial dimensions of the attention map, $\mu_s$ is the mean attention value for the $s$-th attention head. $z_{Bt}$ represents the latent benign text vector, which provides semantic guidance to the attention mechanism.

To ensure perceptual quality while achieving the adversarial objective, we add reconstruction loss to guide the optimization. It enforces latent-level consistency, promote smoothness preserving both low-level details and high-level features between the original and adversarial images. This is mathematically represented as:
\begin{multline}
\mathcal L_{\mathrm{Recon}}
= \frac{1}{H_z \times W_z}
\sum_{i=1}^{H_z}\sum_{j=1}^{W_z}
[\bigl(z_{Bi}^{0}(i,j)-z_{Bi}^{t}(i,j)\bigr)^2 \\[-2pt]
+ 
  \bigl(z_{Bi}^{t}(i+1,j)-z_{Bi}^{t}(i,j)\bigr)^2
  + \bigl(z_{Bi}^{t}(i,j+1)-z_{Bi}^{t}(i,j)\bigr)^2
]
\end{multline}

where $H_z$ and $W_z$ denote the height and width of the latent vector, while $i\in\{1,\dots,H_z\}$ and $j\in\{1,\dots,W_z\}$ index its spatial coordinates. The $z_{Bi}^{0}(i,j)$ refers to the latent value at location $(i,j)$ for the original image. Finally, our whole optimization goal could be defined as:
\begin{equation}
    arg min \quad  \kappa_1 \cdot \mathcal{L}_{Attack} + \kappa_2 \cdot \mathcal{L}_{Recon} + \kappa_3 \cdot \mathcal{L}_{Attention}
    \label{finallyloss}
\end{equation}
The outline of DiffHash is shown in Algorithm \ref{algorithm1}.

\begin{algorithm}[!htbp]

\caption{DiffHash Targeted Attack}
\label{algorithm1}
\textbf{Input}: Clean query image $x$, latent benign text $z_{Bt}$, target image $x_t$, latent target text $z_{Tt}$, Diffusion model $\mathcal{M}$, Deep hashing model $\mathcal{H}$, HashAlignNet $f_{\text{HAN}}$, Loss weights: $\kappa_1, \kappa_2, \kappa_3 $, Total inversion steps $T$ , learning rate $\eta$ 

\textbf{Output}: Adversarial image $x^\prime$
\begin{algorithmic}[1] 
\STATE Obtain latent representation from $x_{i}^{t-1}$: 

$z_{Bi}^{0} = Inverse(x_{i}^{t-1}) =  Inverse \circ  \ldots \circ Inverse(x_{i}^{0})$
\STATE \(\bigl(T_F,b_t\bigr)\;\leftarrow\;f_{\mathrm{HAN}}(z_{Tt})\)
\STATE \(\bigl(I_F,b_{Bt}\bigr)\;\leftarrow\;f_{\mathrm{HAN}}(z_{Bt})\)

\FOR{$t = 1$ to $T$}
\STATE \(z_{Bi}^t \leftarrow \mathrm{diffusion\_step}\bigl(\mathcal M,\;z_{Bi}^{t-1},\;z_{Bt},\;t\bigr)\)  
\STATE 

$    
    \arg \min \; \mathcal{L}_{\mathrm{Total}} = \kappa_1 \cdot \mathcal{L}_{\mathrm{Attack}}(b_{t}, b(z_{Bi}^{t})) + \kappa_2 \cdot \mathcal{L}_{\mathrm{Recon}}(z_{Bi}^{0},z_{Bi}^{t}) + \kappa_3 \cdot \mathcal{L}_{\mathrm{Attention}}(z_{Bt}).
    $

where $b(z_{Bi}^{t})=f_{\text{HAN}}(z_{Bi}^{t})$.

\STATE $        
        z_{Bi}^{t} 
        \;\longleftarrow\; 
        z_{Bi}^{t} - \eta\,\nabla\bigl(\mathcal{L}_{\mathrm{Total}}(z_{Bi}^t,z_{Bt}, z_{Bi}^{0})\bigr)
$

\ENDFOR

\STATE \(x'\;\leftarrow\;{\mathrm{Decode}}(\mathcal{M}, z_{Bi}^T)\)
\STATE \textbf{Output:} Adversarial image $x^\prime$.
\end{algorithmic}
\end{algorithm}
\vspace{-10pt}

\section{Experiment}

\subsection{Experiment Setup and Preparation} \label{4.1}
Following \cite{zhao2023precise}, we evaluate our method on three multi-label datasets: FLICKR-25K (38 labels, 1,900 query/target images)\cite{huiskes2008mir}, NUS-WIDE (21 labels, 2,100 query/target images)\cite{chua2009nus}, and MS-COCO (80 labels, 1,800 query/target images) \cite{lin2014microsoft}, each with 5,000 images randomly selected for hash alignment. For consistency, we use an LLM (e.g., Fuyu-8B \cite{bavishi2023fuyu} using the prompt: \textit{Write a simple five-sentence description of this image.} to generate five descriptions per image, filtering out descriptions with low CLIP similarity ($<$ 0.25) \cite{radford2021learning}. We evaluate targeted attack effectiveness using target mean average precision (t-MAP) on the top 5,000 retrieved images from the database following \cite{zhao2023precise}. Higher t-MAP scores indicate better attack performance.

\textbf{Baselines.} Following \cite{zhao2023precise}, we use CSQ \cite{yuan2020central} with a ResNet-50 backbone as the baseline hashing model. We compare our DiffHash against SOTA targeted attack methods: P2P \cite{bai2020targeted}, DHTA \cite{bai2020targeted}, PTA \cite{zhao2023precise}, Pros-Gan \cite{wang2021prototype}, THA \cite{wang2021targeted}, and HUANG \cite{huang2025huang} , with perturbation constraint $\epsilon=8/255$.

\textbf{Implementation Details.} We employ pre-trained \textit{Stable Diffusion v2-base} with DDIM sampling \cite{songdenoising}. Optimization uses AdamW ($lr=1e^{-3}$, 30 iterations) \cite{loshchilov2017decoupled}. Experiments are conducted on an NVIDIA RTX 4090D GPU, where DiffHash generates adversarial examples with 16.59 GB of GPU memory.

 \begin{table*}[!htbp]

\centering
\begin{tabular}{lcccc ccc ccc} 
\toprule
\multirow{2}{*}{Method} & \multirow{2}{*}{Metric} & \multicolumn{3}{c}{NUS-WIDE} & \multicolumn{3}{c}{FLICKR-25K} & \multicolumn{3}{c}{MS-COCO} \\
\cmidrule(lr){3-5} \cmidrule(lr){6-8} \cmidrule(lr){9-11}
& & \textbf{DPH} & \textbf{HashNet} & \textbf{DPSH} & \textbf{DPH} & \textbf{HashNet} & \textbf{DPSH} & \textbf{DPH} & \textbf{HashNet} & \textbf{DPSH} \\
\midrule
Original& t-MAP & 54.03 & 61.98 & 65.34 & 38.81 & 43.06 & 47.70 & 17.35 & 24.11 & 29.59 \\
P2P& t-MAP & 46.11 & 53.30 & 56.33& 35.64 & 38.94 & 47.50 & 14.65 & 24.39 & 31.53 \\
DHTA& t-MAP & 56.76 & 54.17& 57.43& 40.33 & 43.83 & 52.01 & 15.24 & 26.23 & 34.38 \\
ProS-GAN& t-MAP & 60.55 & \underline{63.77}& 66.37& 39.19 & 43.72 & 44.66 & 22.53 & 23.76 & 39.79 \\
THA& t-MAP & 60.78 & 63.41& \textbf{71.25}& 35.22 & 40.34 & 38.72 & 19.48 & 23.97 & 28.52 \\
PTA& t-MAP & \underline{60.91}& 60.44& 60.85& 46.58& 52.13& \textbf{62.55} & 24.43& 29.26& 42.05\\
 HUANG& t-MAP& 60.05& 56.11& 63.49& \underline{55.26}& \underline{54.61}& 56.94& \underline{43.11}& \underline{42.91}&\underline{49.50}\\
\midrule
\textbf{DiffHash} & t-MAP & \textbf{61.11} & \textbf{64.21}& \underline{67.19}& \textbf{57.04} & \textbf{60.49} & \underline{59.09}& \textbf{60.17} & \textbf{53.39} & \textbf{52.22} \\
\bottomrule
\end{tabular}

\caption{Performance comparison of different methods across three datasets using three hashing models (DPH, HashNet, DPSH) with ResNet-50 backbone and 32-bit hash length.}
\label{tab:cross_hash_performance}
\vspace{-10pt}
\end{table*}

\subsection{Results and Analysis}

We evaluated the t-MAP@5000 of DiffHash against six SOTA methods across three datasets (Table \ref{tab:results1}). DiffHash consistently achieves competitive or superior performance compared to baselines. On MS-COCO, DiffHash notably surpasses ProS-GAN by 27.79\% (16-bit) and PTA by 6.66\% (32-bit). On FLICKR-25K, it closely following THA by only 1.37\% at 16-bit and outperforming baselines in longer hash lengths. On NUS-WIDE, DiffHash particularly excels at 32-bit and 64-bit settings by surpassing PTA and ProS-GAN, although it slightly trails PTA by 2.18\% at the 16-bit setting. Compared to another diffusion-based method HUANG \cite{huang2025huang}, DiffHash achieves improvements ranging from 10.93\% to 21.36\% on NUS-WIDE, 1.52\% to 5.10\% on FLICKR-25K, and 11.05\% to 14.30\% on MS-COCO.

The improved performance of DiffHash mainly comes from its text-guided optimization, which effectively captures more semantic details than label-based methods, especially helpful for large-category datasets like MS-COCO. However, slightly weaker performance observed in shorter hash codes (16-bit on NUS-WIDE and FLICKR-25K) reflects the inherent difficulty in retaining detailed semantic information within limited hash lengths.
\begin{figure}[!htbp]
  \centering
  \includegraphics[width=1 \linewidth ]{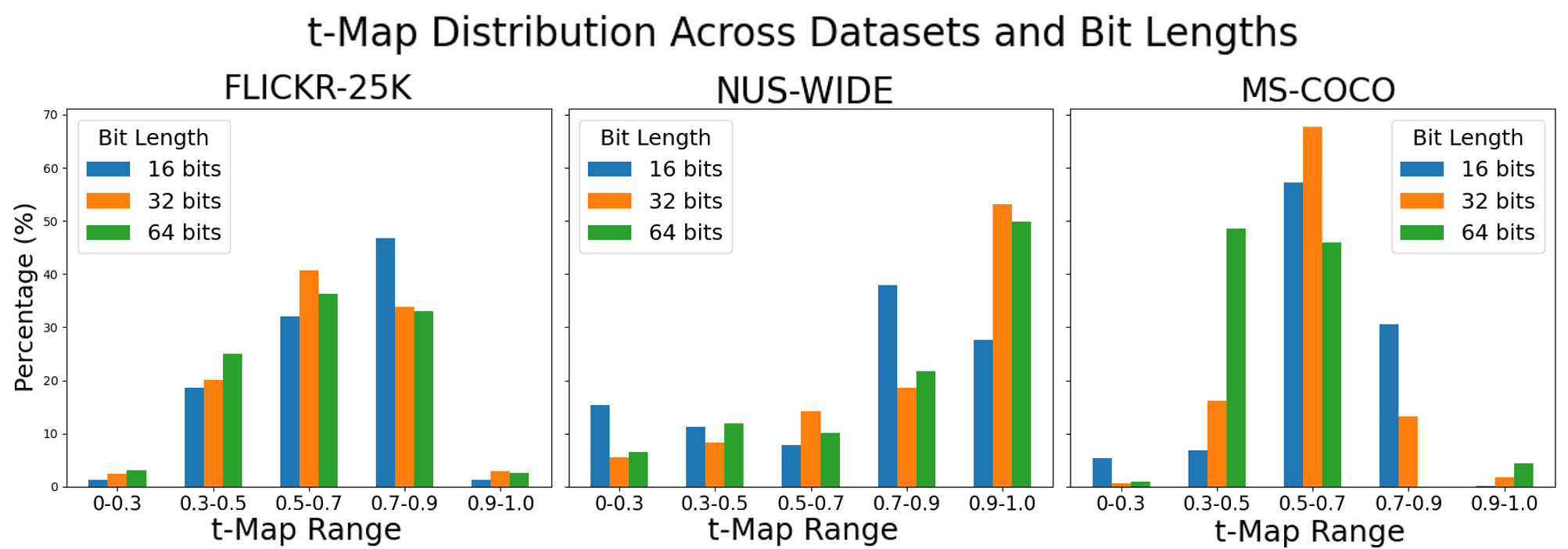}
  \caption{The distribution of t-MAP across three datasets and hash bit lengths (16, 32, 64). }
\label{fig:figure5}
\end{figure}

 \begin{figure*}[!htbp]
  \centering
  \includegraphics[width=0.8 \linewidth ]{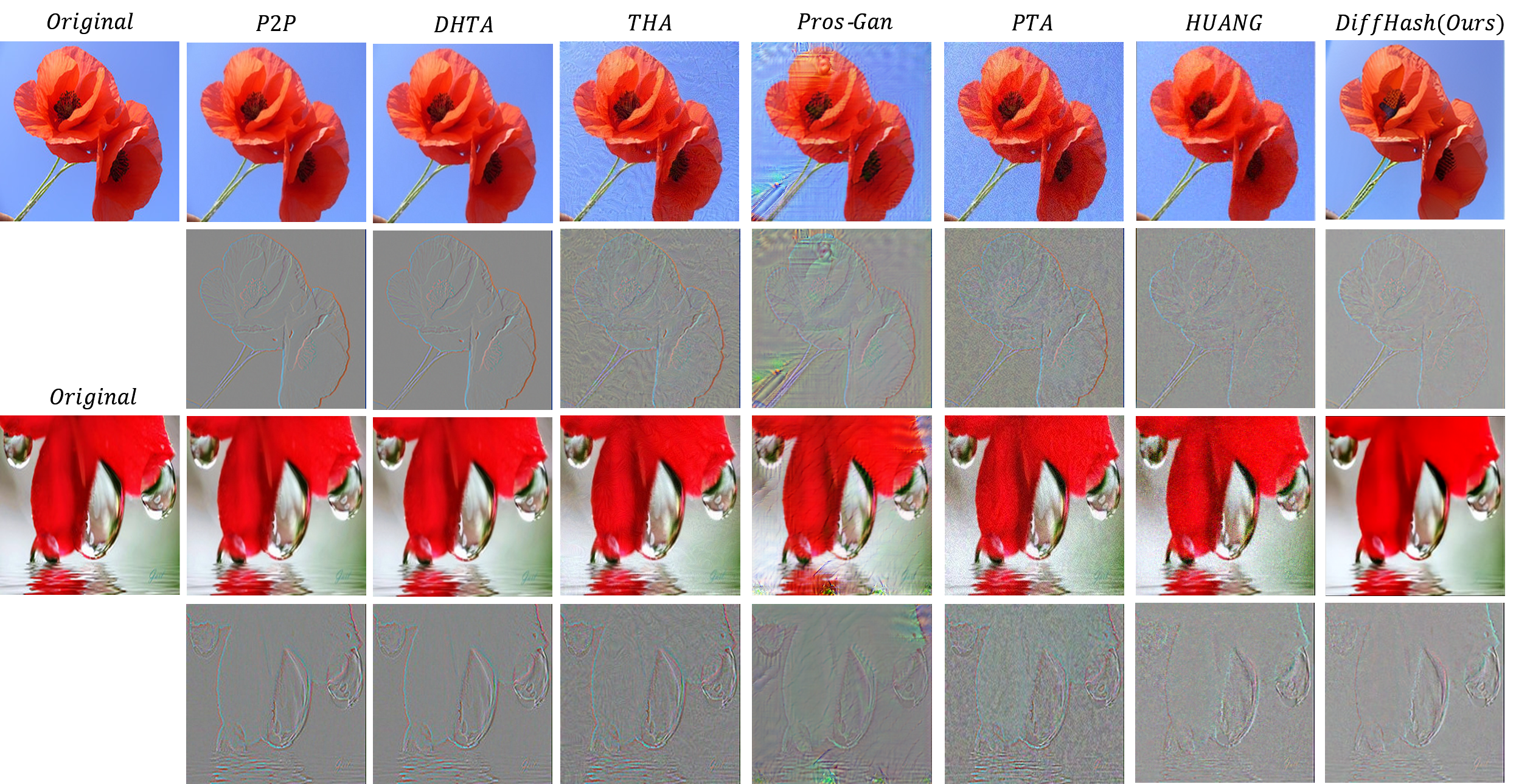}
  \caption{Visual comparison of adversarial examples and perturbation generated by different methods in NUS-WIDE under the same baseline.}
\label{fig:figure6}
\end{figure*}

Figure  \ref{fig:figure5} shows that DiffHash yields consistently high t-MAP scores. On FLICKR-25K and MS-COCO most values lie between 0.5 and 0.9, while on NUS-WIDE over 80 \% of queries exceed 0.9, underscoring its stable and superior targeted‐attack performance across bit lengths. 
\subsection{Universality Across Hashing Methods}

 As shown in Table \ref{tab:cross_hash_performance}, we evaluated the generalization of DiffHash using 32-bit hash codes across multiple hashing methods (DPH, DPSH, HashNet) with a ResNet-50 backbone. DiffHash consistently outperforms SOTA methods across datasets, maintaining approximately 60\% t-MAP across NUS-WIDE, FLICKR-25K, and MS-COCO. As the number of dateset labels increases, methods like THA and ProS-GAN experience significant performance drops, with reductions of over 15\% relative to FLICKR-25K and 30\% compared to NUS-WIDE. Meanwhile, PTA and HUANG also show some decrease which highlights baseline's reliance on label information. Although HUANG performs competitively on MS-COCO, DiffHash consistently outperforms HUANG across all datasets. However, DiffHash slightly lower t-MAP on DPSH for NUS-WIDE and FLICKR-25K can be primarily attributed to DPSH's heavy reliance on pairwise similarity, which limits effective exploitation of the richer semantic consistency available in textual data.

\subsection{Black-box Transferability}

As shown in Table \ref{tab:backbone_performance}, DiffHash demonstrates improved black-box transferability compared to other SOTA methods on other backbones. Our DiffHash outperforms PTA by nearly 19\% and consistently beating HUANG on every architecture. This superior performance can be attributed to the use of rich semantic text information from the target image, in contrast to other approaches that rely solely on label information.
\begin{table}[!htbp]
\centering
\setlength{\tabcolsep}{0.5pt}
\begin{adjustbox}{valign=c}
\begin{tabular}{lccccc}
\toprule
\multirow{2}{*}{Method} & \multicolumn{5}{c}{Backbone Networks} \\
\cmidrule(lr){2-6} 
& ResNet-18 & ResNet-101 & VGG-11 & VGG-16 & AlexNet \\
\midrule
Clean & 8.55 & 7.93 & 7.71 & 6.74 & 8.65 \\
P2P & 10.13 & 9.69 & 9.79 & 9.43 & 10.20 \\
DHTA & 11.09 & 10.57 & 12.16 & 10.70 & 12.11 \\
ProS-GAN & \underline{49.12} & 45.23 & 42.21 & 45.24 & 43.36 \\
THA & 38.95 & 36.64 & 33.34 & 37.40 & 35.20 \\
PTA & 48.58 & \underline{46.46} & 45.34 & 45.48 & 42.19 \\
HUANG & 34.13 & 42.66 & \underline{51.37} & \underline{49.28} & \underline{50.46} \\
\midrule
\textbf{Ours} & \textbf{65.34} & \textbf{65.11} & \textbf{67.34} & \textbf{61.13} & \textbf{62.24} \\
\bottomrule
\end{tabular}
\end{adjustbox}

\caption{Performance comparison of different methods across five backbone networks.}
\label{tab:backbone_performance}
\end{table}

 \subsection{Time Efficiency Analysis and Visual Comparison}

Regarding efficiency, we compared DiffHash with generative attack methods on the NUS-WIDE dataset  as shown in Table \ref{tab:time_efficiency}. DiffHash significantly outperforms GAN-based methods like ProS-GAN and THA by eliminating the need for extensive training. Moreover, operating in the latent space, DiffHash achieves faster processing per image compared to HUANG.
\begin{table}[!htbp]
\centering
  \begin{tabular}{ccl}
    \toprule
    Method & Training Time (s) & Per(s)\\
    \midrule
    ProS-GAN& 36,031 & 0.06  \\
    THA& 35,544 & 0.05  \\
    HUANG& - & 30.11\\
    Ours& -& 19.52\\
    \bottomrule
  \end{tabular}
  \caption{Time efficiency comparison of different generative methods on NUS-WIDE.}
  \label{tab:time_efficiency}
  \vspace{-10pt}
\end{table}

As shown in Figure \ref{fig:figure6}, to evaluate the imperceptibility of DiffHash, we visually compared adversarial examples generated by DiffHash and other SOTA methods. Although methods like DHTA and P2P generate subtle perturbations, their t-MAP remains relatively low. Generative methods like ProS-GAN and THA achieve higher t-MAP but produce noticeable artifacts and visible perturbations. PTA similarly introduces obvious perturbations, whereas HUANG produces smoother images but still displays distortion and blurring. In contrast, DiffHash performs optimization in a low-dimensional, semantically meaningful latent space, enabling highly precise and subtle perturbations, clearly demonstrated in the perturbation visualizations. Thus, DiffHash achieves superior imperceptibility while maintaining effective attack performance.
\begin{table}[!htbp]
\centering
\vspace{-10pt}
\begin{tabular}{lccc c}
\toprule
\multirow{2}{*}{Method} & \multicolumn{3}{c}{Hyperparameters} & \multirow{2}{*}{t-MAP (\%)} \\
\cmidrule(lr){2-4}
 & $\kappa_1$ & $\kappa_2$ & $\kappa_3$ & \\
\midrule
$\kappa_1$ Only         & 15 & 0 & 0 & 42.35 \\
 $\kappa_1$ Only w/o HAN& 15& 0& 0&12.63\\
  \bottomrule
$\kappa_2$ Only         & 0 & 1 & 0 & 1.67 \\
 $\kappa_2$ Only w/o HAN& 0& 1& 0&0\\
  \bottomrule
$\kappa_3$ Only         & 0 & 0 & 8 & 1.67 \\
 $\kappa_3$ Only w/o HAN& 0& 0& 8&0\\
  \bottomrule
$\kappa_1$ and $\kappa_2$ & 15 & 1 & 0 & 59.45 \\
 $\kappa_1$ and $\kappa_2$ w/o HAN& 15& 1& 0&11.79\\
  \bottomrule
$\kappa_2$ and $\kappa_3$ & 0 & 1 & 8 & 0 \\
 $\kappa_2$ and $\kappa_3$ w/o HAN& 0& 1& 8&0\\
  \bottomrule
$\kappa_1$ and $\kappa_3$ & 15 & 0 & 8 & 66.80 \\
 $\kappa_1$ and $\kappa_3$ w/o HAN& 15& 0& 8&14.11\\
\midrule
\textbf{$\kappa_1$+$\kappa_2$+$\kappa_3$}& 15 & 1 & 8 & \textbf{82.10} \\
 \textbf{$\kappa_1$+$\kappa_2$+$\kappa_3$} w/o HAN& 15& 1& 8&22.16\\
 \bottomrule
\end{tabular}
\caption{Ablation study on the influence of different settings on t-MAP.}
\label{tab:ablation_study}

\end{table}

 \begin{figure}[!htbp]
  \centering
  \includegraphics[width=1 \linewidth ]{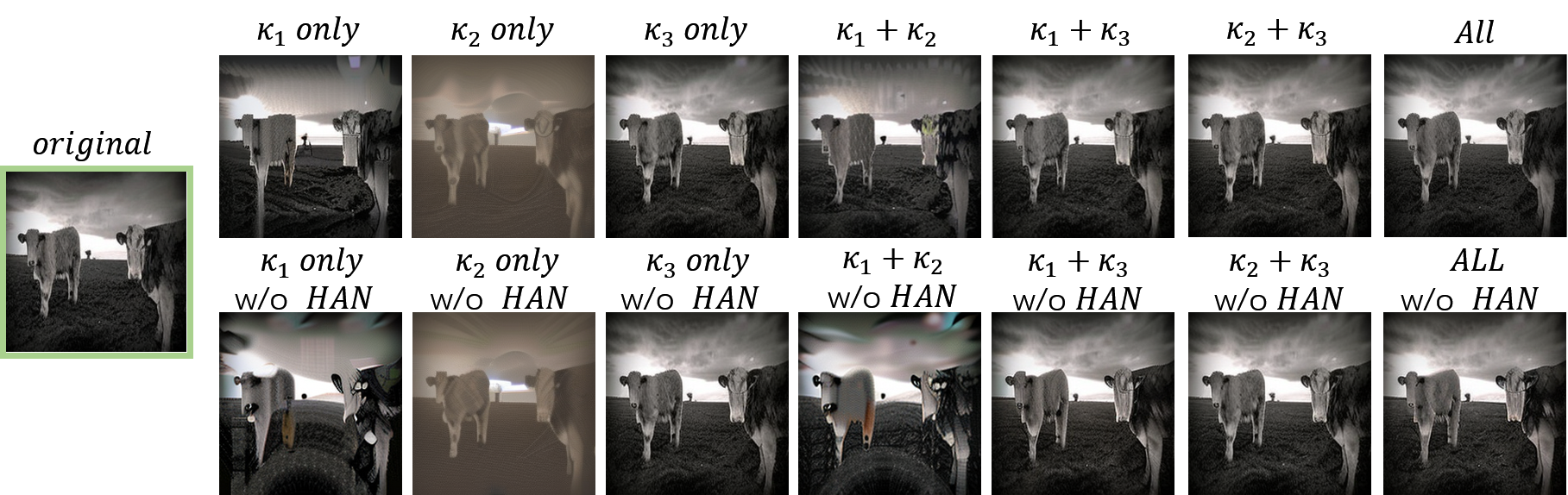}
  \caption{Visual comparison of adversarial examples under different hyperparameter settings.}
\label{fig:figure7}
\vspace{-10pt}
\end{figure}

\subsection{Ablation Study}

Table \ref{tab:ablation_study} and Figure \ref{fig:figure7} illustrate the effects of different settings and the presence of HAN on attack performance and visual quality. Specifically, w/o HAN means that HAN has been removed. Using individual loss functions alone yields poor results, both quantitatively (low t-MAP scores) and visually (noticeable distortions). Combining two losses improves performance moderately. However, the best results are achieved when integrating all three losses with HAN. Notably, omitting HAN significantly reduces performance across all settings, underscoring its essential role in feature alignment.

\section{Conclusion}

In this work, we proposed DiffHash, a novel approach for generating targeted adversarial attacks on deep hashing models by optimizing in the latent space of diffusion model. Unlike traditional methods that rely on label information, DiffHash leverages text guidance to ensure robust attacks. Additionally, we achieved mult-space hash alignment by HAN. Our approach demonstrates significant improvements in black-box transferability and stability across multiple datasets. Extensive experiments show that DiffHash outperforms SOTA methods in terms of both attack effectiveness and visual quality. We hope this work inspires further research into the security of image retrieval models.


\appendix

\bibliography{aaai2026}

\begin{thebibliography}{44}
\providecommand{\natexlab}[1]{#1}

\bibitem[{AI(2023)}]{bavishi2023fuyu}
AI, A. 2023.
\newblock Fuyu-8B: A multimodal architecture for AI agents.

\bibitem[{Bai et~al.(2020)Bai, Chen, Li, Wu, Guo, Xia, and Yang}]{bai2020targeted}
Bai, J.; Chen, B.; Li, Y.; Wu, D.; Guo, W.; Xia, S.-T.; and Yang, E.-H. 2020.
\newblock Targeted Attack for Deep Hashing Based Retrieval.
\newblock In \emph{European Conference on Computer Vision (ECCV)}, 618--634.

\bibitem[{Brown et~al.(2017)Brown, Man{\'e}, Roy, Abadi, and Gilmer}]{advpatch}
Brown, T.~B.; Man{\'e}, D.; Roy, A.; Abadi, M.; and Gilmer, J. 2017.
\newblock Adversarial patch.
\newblock \emph{arXiv preprint arXiv:1712.09665}.

\bibitem[{Cao et~al.(2018{\natexlab{a}})Cao, Long, Liu, and Wang}]{cao2018deep}
Cao, Y.; Long, M.; Liu, B.; and Wang, J. 2018{\natexlab{a}}.
\newblock Deep cauchy hashing for hamming space retrieval.
\newblock In \emph{Proceedings of the IEEE conference on computer vision and pattern recognition (CVPR)}, 1229--1237.

\bibitem[{Cao et~al.(2017)Cao, Long, Wang, and Yu}]{cao2017hashnet}
Cao, Z.; Long, M.; Wang, J.; and Yu, P.~S. 2017.
\newblock Hashnet: Deep learning to hash by continuation.
\newblock In \emph{Proceedings of the IEEE international conference on computer vision (ICCV)}, 5608--5617.

\bibitem[{Cao et~al.(2018{\natexlab{b}})Cao, Sun, Long, Wang, and Yu}]{cao2018deep11}
Cao, Z.; Sun, Z.; Long, M.; Wang, J.; and Yu, P.~S. 2018{\natexlab{b}}.
\newblock Deep priority hashing.
\newblock In \emph{Proceedings of the 26th ACM international conference on Multimedia (ACM MM)}, 1653--1661.

\bibitem[{Chen et~al.(2024)Chen, Chen, Chen, Zhang, Zou, and Shi}]{chen2024diffusion}
Chen, J.; Chen, H.; Chen, K.; Zhang, Y.; Zou, Z.; and Shi, Z. 2024.
\newblock Diffusion models for imperceptible and transferable adversarial attack.
\newblock \emph{IEEE Transactions on Pattern Analysis and Machine Intelligence (TPAMI)}.

\bibitem[{Chua et~al.(2009)Chua, Tang, Hong, Li, Luo, and Zheng}]{chua2009nus}
Chua, T.-S.; Tang, J.; Hong, R.; Li, H.; Luo, Z.; and Zheng, Y. 2009.
\newblock Nus-wide: a real-world web image database from national university of singapore.
\newblock In \emph{Proceedings of the ACM international conference on image and video retrieval (CIVR)}, 1--9.

\bibitem[{Couairon et~al.(2023)Couairon, Verbeek, Schwenk, and Cord}]{couairon2023diffedit}
Couairon, G.; Verbeek, J.; Schwenk, H.; and Cord, M. 2023.
\newblock DiffEdit: Diffusion-based Semantic Image Editing with Mask Guidance.
\newblock In \emph{International Conference on Learning Representations (ICLR)}.

\bibitem[{Dai, Liang, and Xiao(2025)}]{dai2025advdiff}
Dai, X.; Liang, K.; and Xiao, B. 2025.
\newblock Advdiff: Generating unrestricted adversarial examples using diffusion models.
\newblock In \emph{European Conference on Computer Vision (ECCV)}, 93--109. Springer.

\bibitem[{Datar et~al.(2004)Datar, Immorlica, Indyk, and Mirrokni}]{datar2004locality}
Datar, M.; Immorlica, N.; Indyk, P.; and Mirrokni, V.~S. 2004.
\newblock Locality-sensitive hashing scheme based on p-stable distributions.
\newblock In \emph{Proceedings of the twentieth annual symposium on Computational geometry (SoCG)}, 253--262.

\bibitem[{Goodfellow et~al.(2014)Goodfellow, Pouget-Abadie, Mirza, Xu, Warde-Farley, Ozair, Courville, and Bengio}]{goodfellow2014generative}
Goodfellow, I.; Pouget-Abadie, J.; Mirza, M.; Xu, B.; Warde-Farley, D.; Ozair, S.; Courville, A.; and Bengio, Y. 2014.
\newblock Generative adversarial nets.
\newblock \emph{Advances in neural information processing systems (NeurIPS)}.

\bibitem[{Huang and Shen(2025)}]{huang2025huang}
Huang, C.; and Shen, X. 2025.
\newblock HUANG: A Robust Diffusion Model-based Targeted Adversarial Attack Against Deep Hashing Retrieval.
\newblock In \emph{Proceedings of the AAAI Conference on Artificial Intelligence (AAAI)}, 3626--3634.

\bibitem[{Huiskes and Lew(2008)}]{huiskes2008mir}
Huiskes, M.~J.; and Lew, M.~S. 2008.
\newblock The mir flickr retrieval evaluation.
\newblock In \emph{Proceedings of the 1st ACM international conference on Multimedia information retrieval}, 39--43.

\bibitem[{Kingma(2013)}]{kingma2013auto}
Kingma, D.~P. 2013.
\newblock Auto-encoding variational bayes.
\newblock \emph{arXiv preprint arXiv:1312.6114}.

\bibitem[{Kurakin, Goodfellow, and Bengio(2016)}]{IFGSM}
Kurakin, A.; Goodfellow, I.; and Bengio, S. 2016.
\newblock Adversarial machine learning at scale.

\bibitem[{Li et~al.(2021)Li, Gao, Deng, Liu, and Huang}]{li2021adversarial}
Li, C.; Gao, S.; Deng, C.; Liu, W.; and Huang, H. 2021.
\newblock Adversarial attack on deep cross-modal hamming retrieval.
\newblock In \emph{Proceedings of the IEEE/CVF International Conference on Computer Vision(ICCV)}.

\bibitem[{Li et~al.(2022)Li, Lin, Zhou, Qi, Wang, and Jia}]{li2022mat}
Li, W.; Lin, Z.; Zhou, K.; Qi, L.; Wang, Y.; and Jia, J. 2022.
\newblock Mat: Mask-aware transformer for large hole image inpainting.
\newblock In \emph{Proceedings of the IEEE/CVF conference on computer vision and pattern recognition (CVPR)}, 10758--10768.

\bibitem[{Li, Wang, and Kang(2016)}]{li2016feature}
Li, W.-J.; Wang, S.; and Kang, W.-C. 2016.
\newblock Feature learning based deep supervised hashing with pairwise labels.
\newblock In \emph{Proceedings of the Twenty-Fifth International Joint Conference on Artificial Intelligence (IJCAI)}, 1711--1717.

\bibitem[{Lin et~al.(2014)Lin, Maire, Belongie, Hays, Perona, Ramanan, Doll{\'a}r, and Zitnick}]{lin2014microsoft}
Lin, T.-Y.; Maire, M.; Belongie, S.; Hays, J.; Perona, P.; Ramanan, D.; Doll{\'a}r, P.; and Zitnick, C.~L. 2014.
\newblock Microsoft coco: Common objects in context.
\newblock In \emph{European Conference on Computer Vision (ECCV)}, 740--755. Springer.

\bibitem[{Loshchilov(2017)}]{loshchilov2017decoupled}
Loshchilov, I. 2017.
\newblock Decoupled weight decay regularization.
\newblock \emph{arXiv preprint arXiv:1711.05101}.

\bibitem[{Lu et~al.(2021)Lu, Chen, Sun, Wang, Wang, and Yang}]{saa}
Lu, J.; Chen, M.; Sun, Y.; Wang, W.; Wang, Y.; and Yang, X. 2021.
\newblock A smart adversarial attack on deep hashing based image retrieval.
\newblock In \emph{Proceedings of the 2021 international conference on multimedia retrieval (ICMR)}, 227--235.

\bibitem[{Madry(2017)}]{PGD}
Madry, A. 2017.
\newblock Towards deep learning models resistant to adversarial attacks.

\bibitem[{Meng, Chen, and Cao(2024)}]{meng2024targeted}
Meng, F.; Chen, X.; and Cao, Y. 2024.
\newblock Targeted Universal Adversarial Attack on Deep Hash Networks.
\newblock In \emph{Proceedings of the 2024 International Conference on Multimedia Retrieval (ICMR)}, 165--174.

\bibitem[{Mokady et~al.(2023)Mokady, Hertz, Aberman, Pritch, and Cohen-Or}]{mokady2023null}
Mokady, R.; Hertz, A.; Aberman, K.; Pritch, Y.; and Cohen-Or, D. 2023.
\newblock Null-text inversion for editing real images using guided diffusion models.
\newblock In \emph{Proceedings of the IEEE/CVF Conference on Computer Vision and Pattern Recognition (CVPR)}, 6038--6047.

\bibitem[{Moosavi-Dezfooli, Fawzi, and Frossard(2016)}]{deepfool}
Moosavi-Dezfooli, S.-M.; Fawzi, A.; and Frossard, P. 2016.
\newblock Deepfool: a simple and accurate method to fool deep neural networks.
\newblock In \emph{Proceedings of the IEEE conference on computer vision and pattern recognition (CVPR)}, 2574--2582.

\bibitem[{Radford et~al.(2021)Radford, Kim, Hallacy, Ramesh, Goh, Agarwal, Sastry, Askell, Mishkin, Clark et~al.}]{radford2021learning}
Radford, A.; Kim, J.~W.; Hallacy, C.; Ramesh, A.; Goh, G.; Agarwal, S.; Sastry, G.; Askell, A.; Mishkin, P.; Clark, J.; et~al. 2021.
\newblock Learning transferable visual models from natural language supervision.
\newblock In \emph{International conference on machine learning (ICML)}, 8748--8763.

\bibitem[{Rombach et~al.(2022)Rombach, Blattmann, Lorenz, Esser, and Ommer}]{rombach2022high}
Rombach, R.; Blattmann, A.; Lorenz, D.; Esser, P.; and Ommer, B. 2022.
\newblock High-resolution image synthesis with latent diffusion models.
\newblock In \emph{Proceedings of the IEEE/CVF conference on computer vision and pattern recognition (CVPR)}, 10684--10695.

\bibitem[{Saharia et~al.(2022)Saharia, Ho, Chan, Salimans, Fleet, and Norouzi}]{saharia2022image}
Saharia, C.; Ho, J.; Chan, W.; Salimans, T.; Fleet, D.~J.; and Norouzi, M. 2022.
\newblock Image super-resolution via iterative refinement.
\newblock \emph{IEEE transactions on pattern analysis and machine intelligence (TPAMI)}, 4713--4726.

\bibitem[{Shen et~al.(2017)Shen, Gao, Liu, Yang, and Shen}]{shen2017deep}
Shen, F.; Gao, X.; Liu, L.; Yang, Y.; and Shen, H.~T. 2017.
\newblock Deep asymmetric pairwise hashing.
\newblock In \emph{Proceedings of the 25th ACM international conference on Multimedia (ACM MM)}, 1522--1530.

\bibitem[{Song, Meng, and Ermon(2021)}]{songdenoising}
Song, J.; Meng, C.; and Ermon, S. 2021.
\newblock Denoising Diffusion Implicit Models.
\newblock In \emph{International Conference on Learning Representations (ICLR)}.

\bibitem[{Sun et~al.(2024)Sun, Yu, Xie, Li, and Zhang}]{sun2024diffam}
Sun, Y.; Yu, L.; Xie, H.; Li, J.; and Zhang, Y. 2024.
\newblock DiffAM: Diffusion-based Adversarial Makeup Transfer for Facial Privacy Protection.
\newblock In \emph{Proceedings of the IEEE/CVF Conference on Computer Vision and Pattern Recognition (CVPR)}, 24584--24594.

\bibitem[{Szegedy(2013)}]{FGSM}
Szegedy, C. 2013.
\newblock Intriguing properties of neural networks.

\bibitem[{Tang et~al.(2024)Tang, Ye, Lv, Chen, and Zhang}]{tang2024once}
Tang, L.; Ye, D.; Lv, Y.; Chen, C.; and Zhang, Y. 2024.
\newblock Once and for All: Universal Transferable Adversarial Perturbation against Deep Hashing-Based Facial Image Retrieval.
\newblock In \emph{Proceedings of the AAAI Conference on Artificial Intelligence (AAAI)}, 5136--5144.

\bibitem[{Wang et~al.(2017)Wang, Zhang, Sebe, Shen et~al.}]{wang2017survey}
Wang, J.; Zhang, T.; Sebe, N.; Shen, H.~T.; et~al. 2017.
\newblock A survey on learning to hash.
\newblock \emph{IEEE transactions on pattern analysis and machine intelligence (TPAMI)}, 769--790.

\bibitem[{Wang et~al.(2023)Wang, Zhu, Zhang, Zhang, and Han}]{wang2023targeted}
Wang, T.; Zhu, L.; Zhang, Z.; Zhang, H.; and Han, J. 2023.
\newblock Targeted adversarial attack against deep cross-modal hashing retrieval.
\newblock \emph{IEEE Transactions on Circuits and Systems for Video Technology (TCSVT)}, 6159--6172.

\bibitem[{Wang, Lin, and Li(2023)}]{cgat}
Wang, X.; Lin, Y.; and Li, X. 2023.
\newblock Cgat: Center-guided adversarial training for deep hashing-based retrieval.
\newblock In \emph{Proceedings of the ACM web conference (WWW)}, 3268--3277.

\bibitem[{Wang et~al.(2021{\natexlab{a}})Wang, Zhang, Lu, and Xu}]{wang2021targeted}
Wang, X.; Zhang, Z.; Lu, G.; and Xu, Y. 2021{\natexlab{a}}.
\newblock Targeted attack and defense for deep hashing.
\newblock In \emph{Proceedings of the 44th International ACM SIGIR Conference on Research and Development in Information Retrieval (SIGIR)}, 2298--2302.

\bibitem[{Wang et~al.(2021{\natexlab{b}})Wang, Zhang, Wu, Shen, and Lu}]{wang2021prototype}
Wang, X.; Zhang, Z.; Wu, B.; Shen, F.; and Lu, G. 2021{\natexlab{b}}.
\newblock Prototype-supervised adversarial network for targeted attack of deep hashing.
\newblock In \emph{Proceedings of the IEEE/CVF conference on computer vision and pattern recognition (CVPR)}, 16357--16366.

\bibitem[{Xie et~al.(2023)Xie, Zhang, Lin, Hinz, and Zhang}]{xie2023smartbrush}
Xie, S.; Zhang, Z.; Lin, Z.; Hinz, T.; and Zhang, K. 2023.
\newblock Smartbrush: Text and shape guided object inpainting with diffusion model.
\newblock In \emph{Proceedings of the IEEE/CVF Conference on Computer Vision and Pattern Recognition (CVPR)}, 22428--22437.

\bibitem[{Xue et~al.(2023)Xue, Araujo, Hu, and Chen}]{xue2023diffusion}
Xue, H.; Araujo, A.; Hu, B.; and Chen, Y. 2023.
\newblock Diffusion-based adversarial sample generation for improved stealthiness and controllability.
\newblock \emph{Advances in Neural Information Processing Systems (NeurIPS)}, 2894--2921.

\bibitem[{Yang et~al.(2018)Yang, Liu, Deng, and Tao}]{yang2018adversarial}
Yang, E.; Liu, T.; Deng, C.; and Tao, D. 2018.
\newblock Adversarial examples for hamming space search.
\newblock \emph{IEEE transactions on cybernetics}.

\bibitem[{Yuan et~al.(2020)Yuan, Wang, Zhang, Tay, Jie, Liu, and Feng}]{yuan2020central}
Yuan, L.; Wang, T.; Zhang, X.; Tay, F.~E.; Jie, Z.; Liu, W.; and Feng, J. 2020.
\newblock Central similarity quantization for efficient image and video retrieval.
\newblock In \emph{Proceedings of the IEEE/CVF conference on computer vision and pattern recognition (CVPR)}, 3083--3092.

\bibitem[{Zhao et~al.(2023)Zhao, Song, Yuan, Gao, Yang, and Shen}]{zhao2023precise}
Zhao, W.; Song, J.; Yuan, S.; Gao, L.; Yang, Y.; and Shen, H. 2023.
\newblock Precise Target-Oriented Attack against Deep Hashing-based Retrieval.
\newblock In \emph{Proceedings of the 31st ACM International Conference on Multimedia (ACM MM)}, 6379--6389.

\end{thebibliography}

\end{document}